\documentclass[12pt]{iopart}

\expandafter\let\csname equation*\endcsname\relax
\expandafter\let\csname endequation*\endcsname\relax

\usepackage{amsmath,amsthm,latexsym,amssymb,amsfonts,epsfig,bm,graphicx,texdraw,color}
\usepackage[english]{babel}
\usepackage[utf8]{inputenc}

\begin{document}

% repeatedly used symbols shorthands

\newcommand{\HRule}{\rule{\linewidth}{0.4mm}} 
\newcommand{\pd}[2]{\frac{\partial #1}{\partial #2}} 
\newcommand{\td}[2]{\frac{\mathrm{d} #1}{\mathrm{d} #2}} 
\newcommand{\avg}[1]{\left\langle{#1}\right\rangle}
\newcommand{\ex}{\hat{\mathbf{e}}_x}
\newcommand{\ey}{\hat{\mathbf{e}}_y}
\newcommand{\ez}{\hat{\mathbf{e}}_z}
\newcommand{\Rx}{\hat{R}_x}
\newcommand{\Ry}{\hat{R}_y}
\newcommand{\Rz}{\hat{R}_z}
\newcommand{\bzz}{\bm{\zeta}}
\newcommand{\ztt}{\bm{\zeta}^{tt}}
\newcommand{\zrt}{\bm{\zeta}^{rt}}
\newcommand{\ztr}{\bm{\zeta}^{tr}}
\newcommand{\zrr}{\bm{\zeta}^{rr}}
\newcommand{\mrt}{\bm{\mu}^{rt}}
\newcommand{\mtr}{\bm{\mu}^{tr}}
\newcommand{\mrr}{\bm{\mu}^{rr}}
\newcommand{\mi}{\bm{\mu}}
\newcommand{\integr}[3]{\int_{#1}^{#2} \!\! \mathrm{d} {#3}\,}
\newcommand{\ointegr}[3]{\oint_{#1}^{#2} \!\! \mathrm{d} {#3}\,}
\newcommand{\de}{\mathrm{d}}
\newcommand{\oOmega}{\bm{\Omega}}
\newcommand{\bU}{{\bf U}}
\newcommand{\Nabla}{{\nabla}}
\newcommand{\bF}{\mbox{\boldmath $F$}}
\newcommand{\tred}{\textcolor{red}}
\newcommand{\tblu}{\textcolor{blue}}

\newcommand{\sth}{\text{s}\,\theta}
\newcommand{\cth}{\text{c}\,\theta}

\title{Hydrodynamic effects in the capture of rod-like molecules by a nanopore}

\author{Radost Waszkiewicz}
\address{Institute of Theoretical Physics, Faculty of Physics, University of Warsaw}
\ead{rwaszkiewicz@fuw.edu.pl}

\author{Maciej Lisicki}
\address{Institute of Theoretical Physics, Faculty of Physics, University of Warsaw}
\ead{mklis@fuw.edu.pl}
\vspace{10pt}
\begin{indented}
\item[]\today
\end{indented}

\begin{abstract}
In the approach of biomolecules to a nanopore, it is essential to capture the effects of hydrodynamic anisotropy of the molecules and the near-wall hydrodynamic interactions which hinder their diffusion. We present a detailed theoretical analysis of the behaviour of a rod-like molecule attracted electrostatically by a charged nanopore. We first estimate the time scales corresponding to Brownian and electrostatic translations and reorientation. We find that Brownian motion becomes negligible at distances within the pore capture radius, and numerically determine the trajectories of the nano-rod in this region to explore the effects of anisotropic mobility. This allows us to determine the range of directions from the pore in which hydrodynamic interactions with the boundary shape the approach dynamics and need to be accounted for in detailed modelling.
\end{abstract}
%\ioptwocol

\submitto{\JPCM}

\section{Introduction}

Nanopore sequencing is now a well-established technique for the determination of structure of biomolecules~\cite{wanunu2012nanopores,branton2008potential}, such as DNA~\cite{henrickson2000driven}, RNA~\cite{Kasianowicz1996characterization}, or proteins~\cite{waduge2017nanopore}. The molecules, which are typically slender filaments, are electrophoretically transported to the nanopore and then translocated through an orifice. The process of passage or threading, controlled by a combination of electric~\cite{meller2001voltage,forrey2007langevin}, electrokinetic~\cite{ai2011electrokinetic}, entropic~\cite{muthukumar2010}, osmotic~\cite{hatlo2011translocation,jeon2014polymer} and mechanical forces~\cite{szymczak2014}, is now well understood and explored.

However, the approach to the pore is described in less detail. Available models characterise the dynamics of the molecule by its diffusion coefficient $D$ and electrophoretic mobility $\mu_e$. This simplified approach has proved useful to establish the general properties of the system. In Ref.~\cite{grosberg2010dna}, Grosberg and Rabin determined the concentration of DNA near the pore using the Smoluchowski equation formalism. Qiao {\it et al.} defined the capture radius of the pore~\cite{qiao2019voltage}, being the distance at which thermal fluctuations become comparable to the electrostatic potential energy, which bounds the region of attraction of the pore. In the following works, they extended this notion by introducing the orientational capture radius~\cite{qiao2020capture}, being the range at which the electric field strongly orients the colloidal rods. These models, however, neglect the anisotropy of the particles, and of the hydrodynamic interactions with the wall which hinder diffusion at close distances. Hydrodynamic effects are known to alter the trajectories of close sedimenting particles~\cite{Russel1977,Mitchell2015} by coupling to their inherent shape anisotropy and lead to a general slow-down of colloidal dynamics~\cite{Happel}.

In this contribution, we fill this gap by formulating a detailed theoretical approach which accounts for anisotropic diffusivity of a model nano-rod both due to its non-isotropic shape and due to the particle-wall flow-mediated interactions. We first use this model to determine the time scales corresponding to the subsequent phases of motion of a nano-rod approaching a pore: purely Brownian motion far from the pore, electric field-induced translation and reorientation, and the wall influence region. Then we provide a quantitative insight by solving the equations of motion numerically for a collection of initial positions and orientation. We describe the resulting trajectories and mechanisms shaping the motion. This allows us to determine the wall influence region. 

We focus on microparticles which can be modelled as stiff rods. This is appropriate for biomolecules of length $L$ shorter than their persistence length $L_p$. Examples of such nano-rods include dsDNA shorter than $L_p\approx 50\ \text{nm}$ (or ca. $150bp$) or $fd$-viruses~\cite{dogic2001}, with $L=880\ \text{nm}$ and $L_p = 2.8\ \mu\text{m}$, translocating through solid-state pores~\cite{mcmullen2014}. For longer molecules, the effects of elasticity and changing conformation can lead to coiling~\cite{dehaan2013}  or knotting~\cite{szymczak2014,kumar2019} and must be taken into account for proper modelling.

The paper is organised as follows. First, we describe the general model in Sec.~\ref{sec:model-general}, specifying the form of electrostatic interactions in Sec.~\ref{sec:model-electro} and hydrodynamic interactions in Sec.~\ref{sec:model-hydro}. The following Sec.~\ref{sec:results} presents our results, divided into scaling insights in Sec. \ref{sec:results-scalings}, and a numerical analysis of the trajectories in Sec.~\ref{sec:results-trajectories}. We summarise our conclusions in Sec.~\ref{sec:conclusions}.

%The process of their approach to the pore is governed by a combination of electric and hydrodynamic effects. 

%sedimentation 

\section{Model}\label{sec:model}

\subsection{Motion of the nanorod}\label{sec:model-general}

\begin{figure}
    \centering
    \includegraphics[width=0.5\linewidth]{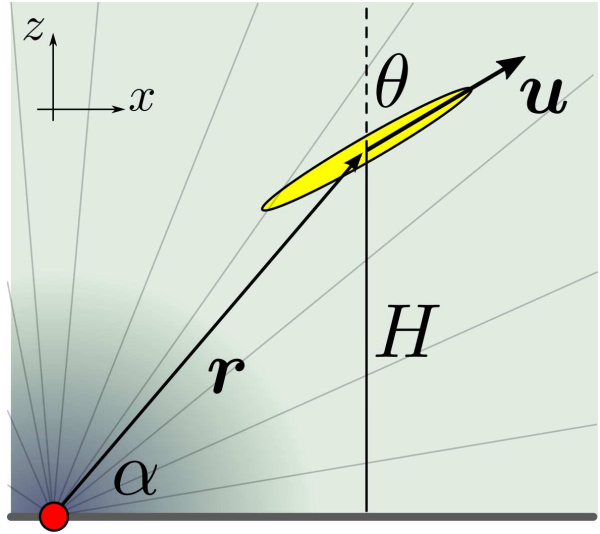}
    \caption{Sketch of a nanorod close to a nanopore denoted by a red dot. The pore generates a radial electric field, modelled by a point charge placed in its location at the origin. The configuration of the rod is given by its position $\bm{r}$ and orientation $\bm{u}$, which corresponds to an inclination angle $\theta$. In addition, we denote by $\alpha$ the polar angle at which the rod is seen from the pore.}
    \label{fig:sketch}
\end{figure}

We consider an ellipsoidal rod of length $L$ with its centre located at a point $\bm{r}$, as sketched in Fig. \ref{fig:sketch}. The aspect ratio of the rod is $p=10$. The director of the rod is a unit vector $\bm{u}$. The nanopore is at the centre of a Cartesian lab coordinate system, with the $z$ axis being normal to the wall, defined by the $xy$ plane. Thus, the rod is at a distance $H = \bm{r}\cdot\bm{e}_z$ from the wall and its inclination angle $\theta$ is determined by $\cos \theta = \bm{u}\cdot\bm{e}_z$.

The rod is charged and moving in an electric field generated by the nanopore. The field exerts an electrostatic force, $\bm{F}_e$ and torque, $\bm{T}_e$, on the rod. On the other hand, the suspending fluid reacts to the motion of the particle by exerting a frictional force, $\bm{F}_h$, and torque, $\bm{T}_h$. Because the flow is characterised by a small Reynolds number, the lack of inertia yields the following equations of motion of the particle
\begin{align}
    \bm{F}_e + \bm{F}_h &= \bm{0},\\
    \bm{T}_e + \bm{T}_h &= \bm{0},
\end{align}
which determine the translational and rotational velocity of the particle, $\bm{V}$ and $\oOmega$, respectively. These are then used to evaluate the trajectory and orientation according to
\begin{align} \label{eq:rdot}
\frac{\de \bm{r}}{\partial t} &= \bm{V}, \\  \label{eq:udot}
\frac{\partial \bm{u}}{\partial t} &= \bm\oOmega \times \bm{u}.
\end{align}

\subsection{Electrostatic interactions with a nanopore}\label{sec:model-electro}

Following previous works~\cite{qiao2020capture}, we model the interaction between a pore and a rod using the Coulomb potential from a point source. We assume that in the capture process rods are uniformly charged with an effective electrophoretic charge $Q$~\cite{grosberg2010dna}. We will measure the strength of the electric field by the capture radius $\lambda_e$: a distance where the potential energy $Q\Psi$ of the rod is comparable to the thermal fluctuations, so $Q\Psi(\lambda_e) = k_B T$, with $k_B$ being the Boltzmann constant and $T$ the temperature. In this setting, the electric field at a location $\bm{r}$ can be written as
\begin{equation}\label{elefield}
 \bm{E}(\bm{r}) = -\frac{k_B T \lambda_e}{Q}\frac{\bm{\hat{r}}}{r^2},
\end{equation}
where the hat denotes a unit vector. The potential energy of the particle in the electric field is given by an integral along the rod
\begin{equation}\label{eleenergy}
    {\Psi}_{e} = \frac{k_B T \lambda_e}{L} \int_{-L/2}^{L/2} \frac{\de s}{r(s)} ,
\end{equation}
where $r(s)$ represents distance from the pore to the rod element $s$. From this we can obtain exact integral expressions for the force $\bm{F}_e$ and torque $\bm{T}_e$ acting on the rod by appropriate differentiation. Since in the remainder of the paper we will be interested in intermediate particle-pore distances, when $L/r \ll 1$, we expand them and to leading order we find
\begin{align} \label{eq:force}
    \bm{F}_e & = - k_B T \lambda_e \frac{\bm{\hat{r}}}{r^2} + \mathcal{O}({(L/r)^{2}}), \\
     \bm{T}_e &= - \frac{k_B T \lambda_e L^2}{4} \frac{(\bm{\hat{r}}.\bm{u})(\bm{\hat{r}}\times\bm{u})}{r^3} + \mathcal{O}({(L/r)^{4}}).
\end{align}
The force is radially attracting the rod and falls off as $r^{-2}$. When rod is oriented at an angle to the direction towards the pore, uneven force distribution generates a torque which reorients the rod, forcing it to point towards the pore. This torque is proportional to the electric field gradient, thus it scales as $r^{-3}$.

\subsection{Near-wall mobility of a rod-like particle}\label{sec:model-hydro}

On the colloidal length and time scales, relevant for nanopore experiments, the flow field $\bm{v}(\bm{r})$ around a particle is described by the stationary Stokes equations~\cite{KimKarrila}
\begin{equation}\label{stokes}
 \eta{\nabla}^2\bm{v}(\bm{r}) -{\nabla} p(\bm{r}) = - \bm{f}(\bm{r}),\qquad {\nabla}\cdot\bm{v}(\bm{r}) = 0,
\end{equation}
where $\bm{f}(\bm{r})$ {is the force density the particle exerts on the fluid}, and $p(\bm{r})$ is the modified pressure field.
If a particle is moving in a quiescent fluid, the frictional force and torque are linearly related to its translational and angular velocities, $\bm{V}$ and $\bm{\Omega}$, via the friction (or resistance) tensor~\cite{KimKarrila,EkielJezewska2009}
\begin{equation}\label{friction_single}
\begin{pmatrix}
\bm{F}_h \\
\bm{T}_h 
\end{pmatrix} = -  
\begin{pmatrix}
\bzz^{tt} & \bzz^{tr}  \\
\bzz^{rt} & \bzz^{rr}  \\ 
\end{pmatrix}
\begin{pmatrix}
\bm{V}  \\
\oOmega
\end{pmatrix}.
\end{equation} 
 The indices $tt$ and $rr$ above denote the translational and rotational parts, respectively, while the tensors $\bzz^{tr}$ and $\bzz^{rt}$ describe the translation-rotation coupling. 
 
 If a particle is moving under the action of a known force and torque, a complementary problem is formulated using the mobility tensor ${\mi}$  which is an inverse of the friction tensor
\begin{equation} \label{frictionmobility}
\mi = \begin{pmatrix}
\mi^{tt} & \mi^{tr} \\
\mi^{rt} & \mi^{rr} 
\end{pmatrix} = 
\begin{pmatrix}
\bzz^{tt} & \bzz^{tr} \\
\bzz^{rt} & \bzz^{rr} 
\end{pmatrix}^{-1}  = \bzz^{-1}.
\end{equation}
Finally, the mobility tensor is related to the diffusion matrix $\bf{D}$ by the fluctuation-dissipation theorem
\begin{equation}
    \bm{D} = k_B T \mi.
\end{equation}
Thus, the diffusive properties of the particle are completely determined by its hydrodynamic mobility. 

For an axisymmetric particle, the friction (and mobility) tensors have a high degree of symmetry. In a bulk system, the configuration of a particle is given by its axial unit vector $\bm{u}$, and the friction tensor can be explicitly written as
\begin{align} \label{eq:bulktt}
    \bzz^{tt} &= \zeta^{t}_\parallel \bm{u}\bm{u} + \zeta^{t}_\perp (\bm{1}-\bm{u}\bm{u}), \\
    \bzz^{rr} &= \zeta^{r}_\parallel \bm{u}\bm{u} + \zeta^{r}_\perp (\bm{1}-\bm{u}\bm{u}), \\
    \bzz^{tr} &= \bzz^{rt} = \bm{0},
\end{align}
using only four coefficients. For ellipsoids, analytical formulae are available for the bulk diffusion tensor and are listed in \ref{appEllipsoid}. Otherwise, efficient schemes for the calculation of bulk hydrodynamics properties of macromolecules modelled as collections of beads, such as {\sc Hydro++} ~\cite{delatorre2007,fernandes2002brownian}, {\sc GRPY}~\cite{zuk2018grpy}, or {\sc Hydromultipole}~\cite{Cichocki1994g,EkielJezewska2009} are also available.

The presence of a confining boundary changes this situation, since the hydrodynamic tensors now depend both on the distance to the boundary, and on the orientation of the nano-rod with respect to the surface. The  friction tensors of a near-wall particle, $\bzz_w$, may be written as:
 \begin{equation}\label{correctio}
  \bzz_w = \bzz_0 + \bm{\Delta}\bzz_w,
 \end{equation}
 where $\bzz_0$ is the bulk resistance tensor, and the last term is a wall-induced correction. An analytical leading-order approximation to $\bm{\Delta}\bzz_w$, with the expansion parameter $L/H$ being the ratio of the size of the particle, $L$, to the wall-particle distance $H$, was derived previously by some of us~\cite{Lisicki2016nearwall}. Earlier works provide the components of the diffusion tensor for very slender filaments close to walls only for particular alignments~\cite{Katz1975,DeMestre1975}. The treatment proposed in Ref.~\cite{Lisicki2016nearwall} allows for an efficient calculation of the diffusion tensor of a slender rod-like particle for moderate and large wall-particle distances. 
  By inverting $\bzz_w$ from Eq. \eqref{correctio}, we arrive at a convenient approximation to the near-wall mobility $\mi_w = \bzz_w^{-1}$, which will serve as the starting point for present work. The correction terms have the following form
\begin{align}\label{corrtt} 
\bm{\Delta}\bzz^{tt}_w =& -\frac{\mathbf{A}_1}{2H}  
+ \frac{\mathbf{A}_2}{(2H)^2}  +\mathcal{O}(H^{-3}), \\
\bm{\Delta}\bzz^{tr}_w =& -\frac{\mathbf{B}}{(2H)^2}  + \mathcal{O}(H^{-3}), \\
\bm{\Delta}\bzz^{rt}_w =& -\frac{\mathbf{B}^\mathrm{T}}{(2H)^2}  + \mathcal{O}(H^{-3}), \\ \label{corrrr}
\bm{\Delta}\bzz^{rr}_w =& -\frac{\mathbf{C}}{(2H)^3}+ \mathcal{O}(H^{-4}).
\end{align}
The tensors $\mathbf{A}_{1,2}$, $\mathbf{B}$, $\mathbf{C}$ above are derived from the multipole expansion of the Blake tensor~\cite{Blake1971} (Oseen tensor for the wall-bounded geometry) and they depend on the bulk components of the friction tensor of a rod-like particle and its orientation angle $\theta$ but not on the wall-particle distance. For completeness, we write them explicitly in \ref{appA}.

Notably, there are other strategies for tackling the problem of near-wall mobility or various levels of accuracy, such as boundary integral equations~\cite{Hsu1989} or finite element method simulations~\cite{DeCorato2015}. Approximate bead-model numerical schemes on a similar level of accuracy to the analytical correction above supplemented by lubrication treatment of close configurations have been introduced by Swan and Brady~\cite{Swan2007}.  More accurate multipole expansion approaches have also been developed~\cite{Cichocki2000wall}. However, comparisons of the analytical correction presented above in Eqs.~\eqref{corrtt}–\eqref{corrrr} with accurate bead-model {\sc hydromultipole} scheme~\cite{Cichocki2000wall} for a rod of aspect ratio $p=10$ have shown the validity of the correction for distances up to $H/L \sim 1$, provided that the rod is far from touching the wall, in which case lubrication corrections become important. Thus, in this contribution we will use the approximate correction, bearing in mind that the analysed model has been developed for moderate particle-nanopore distances.
 Here, we restrict our attention to the semi-analytical scheme presented above, since the dynamics occur mostly in the range of wide separation of the particle and the wall. 
%As discussed in detail in Ref.~\cite{Lisicki2016nearwall}, the optimal strategy with regard to accuracy is to find the diffusion tensor by first calculating the wall-corrected friction tensor, and then inverting it to obtain the mobility tensor. 

%\subsection{Langevin and Brownian Dynamics simulations}

\section{Results}\label{sec:results}

\subsection{Scaling analysis} \label{sec:results-scalings}

The length of the particle is $L$. For dsDNA the length of $L=100\ bp$ corresponds to ca. $34\ \text{nm}$. Another length scale in the problem is the electrostatic capture radius ${\lambda_e}$, which at room temperature is in the range of micrometers~\cite{qiao2020capture}, so $\lambda_e/L \gg 1$.

\begin{table*}
\caption{\label{tab:timescales}Time scales in the system}
\rm
\begin{tabular*}{0.7\textwidth}{lll}
\br
Symbol &Scaling &Description\\
\mr
$\tau$ & $\displaystyle {6 \pi \eta L^4 }/{k_B T \lambda_e}$ & Basic time scale \\
$\tau_B$ & $\displaystyle {6 \pi \eta L^3}/{k_B T}$ & Brownian (diffusive) time \\
$\tau_e^t$ & $\displaystyle {6\pi \eta L^2 r^2}/{ k_B T \lambda_e}$ & Electrostatic translation time \\
$\tau_e^r$ & $\displaystyle {6\pi \eta L r^3 }/{k_B T \lambda_e} $  & Electrostatic rotation time \\
$\tau_w$ & $ \displaystyle {6 \pi \eta  H^2 r^2 }/{k_B T \lambda_e }$ & Wall-induced rotation time \\
\br
\end{tabular*}
\end{table*}

\begin{figure}
    \centering
    \includegraphics[width=0.8\linewidth]{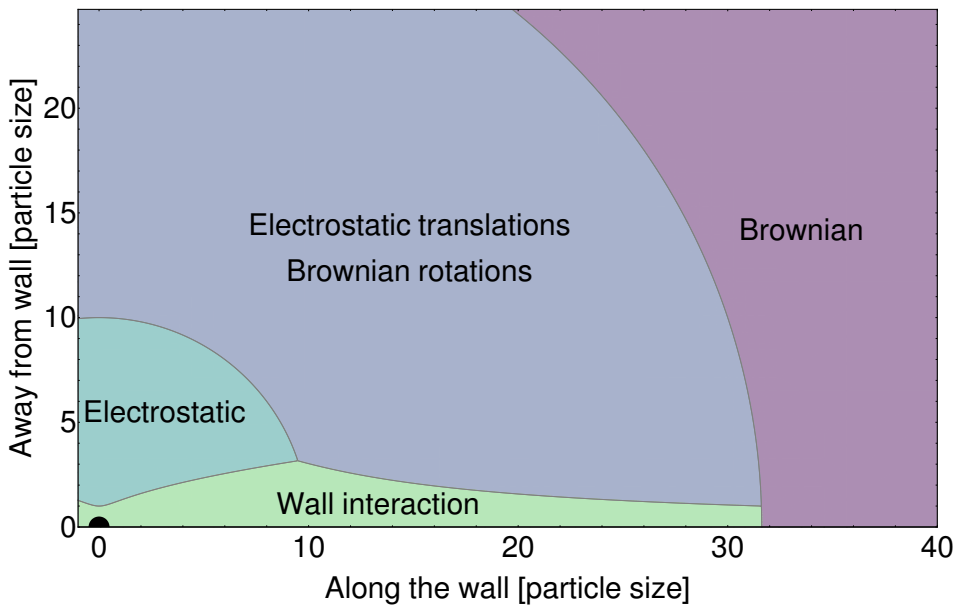}
    \caption{The division of near-pore space into regions colored by dominant terms determining the dynamics of the rod. The pore is located at the origin, and the wall coincides with the bottom border of the graph. Closest to the surface, wall interaction terms are the most important. When the particle is moving further away from the wall, we expect concentric regions of electrostatically determined dynamics, electrostatic translations with Brownian rotations, and fully Brownian dynamics. This corresponds to the intuition that electrostatic torque decays faster than force when moving away from the pore. The boundaries between subsequent regions are obtained by comparing the respective times scales of motion. Expressions used to determine the time scales are collected for convenience in Table \ref{tab:timescales}. For this calculation we assumed $\lambda_e =10^3 L$.}
    \label{fig:regions}
\end{figure}

The interplay between different time scales in the problem determines the ranges in which different mechanisms of capture dominate. We list them all for convenience in Table \ref{tab:timescales}.  Far enough from the pore, the dynamics are purely Brownian, and the relevant time scale is $\tau_B\sim L^2/\langle{D^t}\rangle \approx 6\pi \eta L^3 / k_B T $, where $\avg{D^t}$ is the average diffusion coefficient of the nano-rod. Rotational Brownian motion occurs on the same time scale $1/\avg{D^r}  \sim 8 \pi \eta L^3 / k_B T$. In the presence of an electrostatic force \eqref{eq:force} driving the translational motion, the velocity scales as $V\sim  \avg{D^t} \lambda_e / r^2 \sim k_B T \lambda_e / 6 \pi \eta L r^2$. Thus the relevant translational time scale is $\tau^t_e \sim L/V \approx 6 \pi \eta L^2 r^2/ k_B T \lambda_e$.  The electrostatic torque falls off quicker with distance, so the relevant rotation time scale derived from it becomes $\tau^r_e  \approx 6 \pi \eta L^2 r^3/ k_B T \lambda_e$.  By comparing the Brownian and electrostatic rotational time scale, we recover the scaling for the orientational capture radius defined in Ref.~\cite{qiao2020capture}. We note that in the mobility (or diffusion) matrix $\bm{\mu}$, the translational elements scale as $\eta L$, the coupling terms as $\eta L^2$, and the rotational terms as $\eta L^3$. This scaling changes close to the wall, where an additional length scale, the wall-particle distance $H$, comes into play. When the rod comes close to the wall, reorientation due to hydrodynamic interactions with the boundary becomes important, with the relevant time scale $\tau_w$ derived from the scaling form of the equation $\oOmega = \mi^{rt}\cdot\bm{F}_e + \mi^{rr}\cdot\bm{T}_e $. By comparing different time scales we determine 4 general regimes of motion, sketched in Fig. \ref{fig:regions}. The furthest region is Brownian, but closer to the pore when $\tau_B\sim \tau^{t}_e$, translational motion is driven by electrostatic forces but rotational motion remains Brownian. Moving even closer, electrostatic torque dominates over thermal reorientation. The boundaries between these regions are determined by the radial distance from the pore. However, in the vicinity of the wall, the diffusivity and hence mobility are generally hindered by hydrodynamic interactions with the wall, which become the dominant driving mechanism.
We thus conclude that for the analysis of the behaviour of a field-driven nano-rod, since we focus on the dynamics in the range $L<r\ll\lambda_e$, Brownian motion may be neglected, as it would only influence the initial orientations with which the rod would enter the area dominated by electrostatic interactions. It is essential, however, to retain the hydrodynamic anisotropy of the rod, as it influences both the translational and rotational motion in the presence of a strong electric field.

For quantitative calculations, we choose dimensionless units with the basic length $L$. In the presence of an electric field, the more appropriate time unit is $\tau = \tau_B L/\lambda_e = 6\pi \eta L^4 / k_B T \lambda_e$. For dsDNA mentioned above, this time scale is of the order of $10^{-7}\ \text{s}$. With this choice, we can write the force and torque acting on the rod as
\begin{equation}
    \bm{F}_e = - \frac{\bm{\hat{r}}}{r^2}, \qquad \bm{T}_e = - \frac{(\bm{\hat{r}}\cdot\bm{u})(\bm{\hat{r}}\times\bm{u})}{r^3}.
\end{equation}

\subsection{Discussion of trajectories}\label{sec:results-trajectories}

Before focusing on the numerical solutions of the full set of equations, it is informative to consider a very simple case of an anisotropic particle subject to a central force without the wall influence and with no external torque. In this case, the particle's mobility tensor has the form as in Eq.~\eqref{eq:bulktt}, with two coefficients $\mu_\parallel$, $\mu_\perp$. For very slender rods, we have $\mu_\parallel\approx 2\mu_\perp$.

We can describe such a situation by taking a coordinate system centred at the pore, and oriented with the principal axes of the body ($\parallel,\perp)$. Within this parametrisation, the equations of motion of the centre of the rod $(r_\parallel,r_\perp)$ are
\begin{eqnarray}
 \frac{\partial r_\parallel}{\partial t}  &= \mu_\parallel \frac{F(r)}{r} r_\parallel  \\
  \frac{\partial r_\perp}{\partial t}   &= \mu_\perp \frac{F(r)}{r} r_\perp 
\end{eqnarray}
By displaying the equations in this form, it is immediately clear that the function $F(r)$ only influences the time dependence and has no bearing on the trajectory. For bounded $F$, there is a hyperbolic fixed point at the origin with two important trajectories intersecting at it: $r_\parallel = 0$ and $r_\perp = 0$. The first corresponds to slower, sideways motion and the second to faster, axial motion. We conclude that a particle with different values of the drag coefficients along different axes will almost always approach the fixed point along the slowest axis, which in the case of a rod-like colloids means that the particles would most often approach the pore broadside. This conclusion holds sway whenever the torques acting on the rod are negligible and thus it concisely describes the initial dynamics when starting far enough from the pore. Approaching the slow rectilinear trajectory is accomplished by hydrodynamic 'gliding' effect where velocity stays at an angle to the force in the initial stages of motion. Moreover, we predict the existence of two types of trajectories -- concave and convex -- depending on whether rod initially points towards or away from the wall compared to the pore direction.

\begin{figure}
    \centering
    \includegraphics[width=0.8\linewidth]{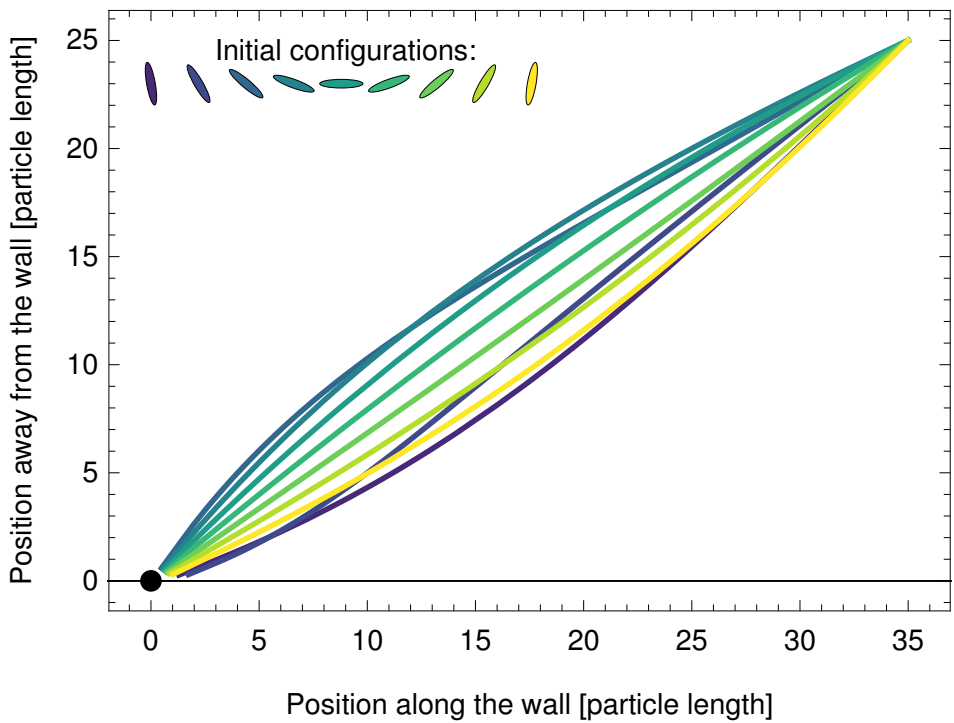}
    \caption{Trajectories of the article centre depending on the initial orientation of the particle. Due to differences of drag coefficients in the direction along and across the particle, the velocity is never aligned with the drag force. This leads to gliding, either higher above the wall or closer to the wall depending on the orientation, importantly ruling out a straight path towards the pore.}
    \label{fig:traj_orient}
\end{figure}

To explore the dynamics driven by the interplay between electrostatic and hydrodynamic forces acting on the nanorod, we integrate the deterministic equations of motion numerically. As argued before, we neglect the influence of Brownian motion in the range of distances under consideration. The equations of motion for the system are
\begin{align} \label{eq:model}
\begin{pmatrix}
\bm{V}  \\
\oOmega
\end{pmatrix} &= \mi_w   \begin{pmatrix}
\bm{F}_e \\
\bm{T}_e 
\end{pmatrix}, 
\end{align}
which together with Eqs. \eqref{eq:rdot} and \eqref{eq:udot} allow us to calculate the trajectories of nano-rods. We consider an ellipsoidal rod of aspect ratio $p=10$. We integrated the equations of motion using \texttt{NDSolve} command of Mathematica 12.1 with \texttt{Method -> \{"EquationSimplification" -> "Residual"\}} option enabled to deal with algebraic-differential nature of the equations.

\begin{figure}
    \centering
    \includegraphics[width=0.8\linewidth]{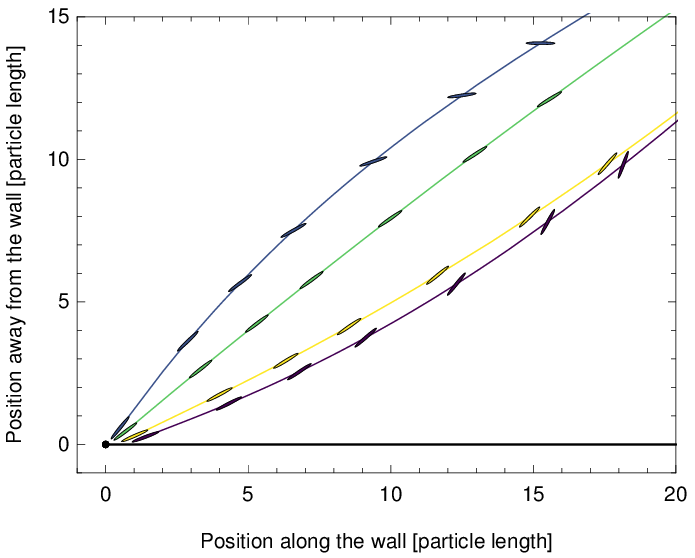}
     \includegraphics[width=0.8\linewidth]{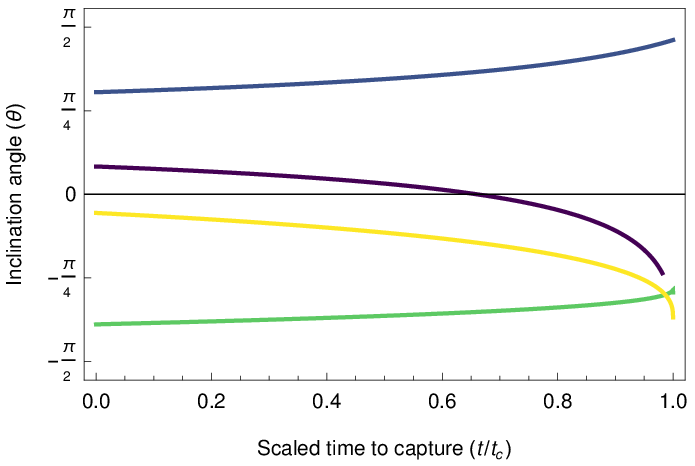}
    \caption{{\bf Top}: Four sample trajectories with the particle orientation sketched at equal translation intervals. The particles start from an initial position $(35,25)$ at four different initial orientations ($\theta_0=k \pi/36$, with $k=3,11,22,34$), depending on which they follow convex or concave gliding trajectories.
    {\bf Bottom}: Inclination angles corresponding to the trajectories plotted above against normalized time. Reorientation rapidly accelerates in the later stages of motion because the initial offset due to the sideways glide close to the starting point has increasingly larger effect on the approach angle.    }
    \label{fig:traj_four}
\end{figure}

We present the resulting trajectories in Fig.~\ref{fig:traj_orient}, starting from a point $\bm{r}_0= (35,25)$ above the wall, thus seen at an angle $\alpha_0 \approx 2\pi/9$ from the nanopore. Rods starting at different orientations glide sideways with respect to the field direction due to their shape (and therefore also mobility) anisotropy. For the initial angle $\theta_0<\alpha_0$, the trajectories are convex and they approach the vicinity of the pore from below the direction $\alpha_0=\theta_0$. The initial inclination $\theta_0>\alpha_0$ leads to a concave path gliding above the $\bm{r}_0$ direction. The shape of the trajectories is also dependent on the aspect ratio of the nano-rod, which determines the parallel and perpendicular friction (mobility) coefficients. At large distances, the wall does not influence the observed dynamics. A closer look into the particles' orientations, sketched in Fig.~\ref{fig:traj_four} for four chosen initial orientations ($\theta_0=k \pi/36$, with $k=3,11,22,34$), reveals a strong alignment with the field lines, predicted by earlier works neglecting hydrodynamic anisotropy~\cite{qiao2020capture}. The time dependence of the rods' orientation shows a systematic change, the rate of which increases when approaching the pore due to the increasingly strong attractive force and aligning torque. At the considered separations from the pore, we see no pronounced effect of the translation-rotation coupling due to the wall, which in this case is too weak to compete with field-driven motion. It would come into significance at near-touching configurations where we also expect the detailed geometry of the pore to matter.

\begin{figure}
    \centering
    \includegraphics[width=0.8\linewidth]{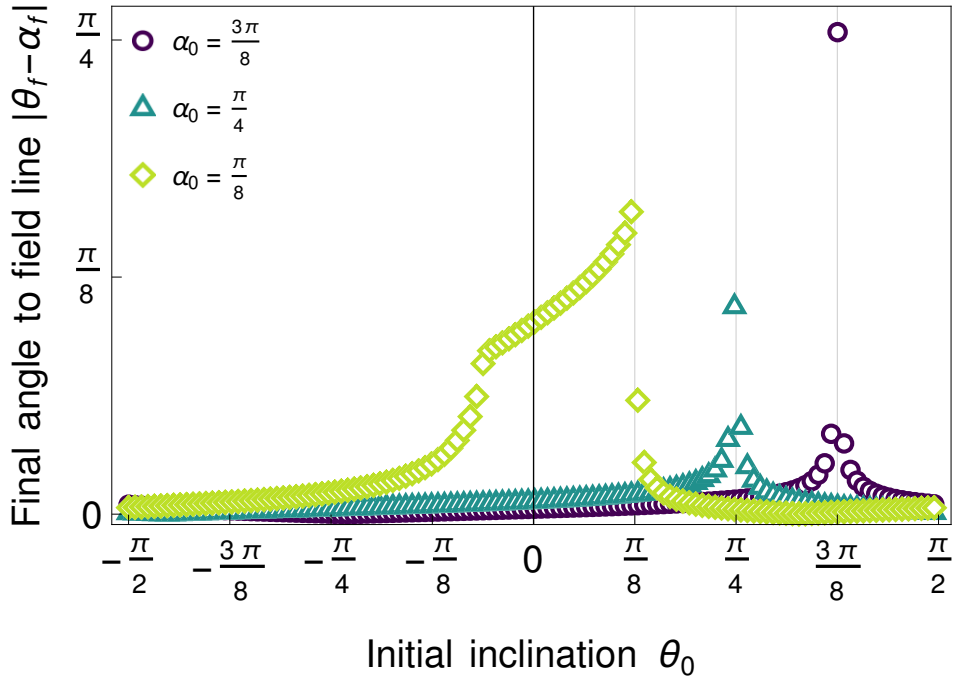}
    \caption{The final angle between the rod and the field line plotted against the initial inclination angle at different starting positions far away from the pore with different polar angles: $\alpha_0 = \pi/8$ -- close to the wall, $\alpha_0 = \pi /4$ -- intermediate, $\alpha_0 = 3\pi/8$ -- far from the wall. Trajectories with $\theta_0 < \alpha_0$ are convex and initially glide towards the wall, and trajectories with $\theta_0 > \alpha_0$ are concave and initially glide away from the wall. Importantly, when the two angles are similar, $\theta_0\approx\alpha_o$, a straight trajectory is unstable because of drag anisotropy and leads to relatively large values of the final angle between the field lines and the rod, $|\theta-\alpha|$. Additionally, for small $\alpha_0$ this relationship is asymmetric due to wall influence where the hydrodynamic torque from wall drag competes with electrostatic reorientation in the late stages of movement.}
    \label{fig:final_angle}
\end{figure}

To test how strong the alignment with the field lines is, in Fig.~\ref{fig:final_angle} we present results of numerical simulations for a spectrum of initial orientation angles $\theta_0$ for rods released far away from the pore at three representative values of the polar angle $\alpha_0$. For each starting configuration, we determine the final orientation $\theta_f$ and plot the final angle to the local field line,  $|\alpha_f-\theta_f|$, at the final position of the particle. For most initial values of the approach angle, conclusions are similar to those from models neglecting wall interaction terms - the final deviation angle from the field lines direction is typically of the order of $\pi/50$. For small angles $\alpha_0$, however, the relationship between the initial inclination and  the final angle to field lined is asymmetric due to wall interactions, which increase the offset to the field lines. For example, we observe angles greater than $\pi/16$ in the region $(-\pi/8, \pi/8)$. When interpreting this result, it is notable that in 3D under uniform distribution of charged particles over the hemisphere of possible initial directions, small angles $\alpha_0$ occur much more often (proportionally to $\cos\alpha_0$), meaning that such effect can hinder the capture of a substantial number of particles. The area of the region with $\alpha_0<\pi/8$ accounts for nearly 40\% of the considered hemisphere. Thus it remains important in a high percentage of capture events to properly resolve hydrodynamic interactions with the confining boundary.

Notably, we have chosen the initial orientations to lie within the $xz$ plane. By solving the full, three-dimensional equations of motion~\eqref{eq:rdot}, \eqref{eq:udot} and \eqref{eq:model}, we confirm that the qualitative characteristics of motion remain unchanged even when the initial orientation of the rod has an out-of-plane component. The reorientation in the region far away from the wall remains mostly Brownian, while in the near-pore region the strong alignment mechanism brings the dynamics to a plane to which we restrict our attention from the beginning. We thus do not see any changes as compared to the axisymmetric configurations analysed here.

\section{Conclusions}\label{sec:conclusions}

We have presented an analysis of motion of a colloidal nano-rod driven by the electric field of a nanopore in a viscous fluid with a particular focus on the inclusion of hydrodynamic interactions and a detailed analysis of the different regimes of motion. The nanopore is modelled as a point charge which attracts a uniformly charged rod-like particle. Basing on scaling arguments, we identified the various time scales of motion and demonstrated that far away from the pore the motion of the particle is purely Brownian but as soon as it reaches the electrostatic capture radius $\lambda_e$ it is systematically attracted towards the pore. Its initial dynamics are then governed by an electrostatic force driving its translational motion, with the velocity resulting from the balance of this force and the fluid drag force. Since the latter is anisotropic, and depends on the orientation of the particle, the motion resembles sideways gliding towards the pore.  At closer distances, the electrostatic torque becomes pronounced and strongly aligns the rods with the electric field lines, as reported previously~\cite{qiao2020capture}. However, earlier studies neglected the role of hydrodynamic interactions both on the level of the particle anisotropy, and the wall-induced increase of friction. 

In this contribution, we have outlined a theoretical approach which takes into account both the anisotropy of the particle and the wall hindrance effect. Supported by scalings, we explored the trajectories at intermediate distances, when Brownian motion can be neglected, but the rod is far enough from the pore to disregard the field and flow effects of the pore geometry. At large and moderate distances, we have employed an approximate analytical scheme, in which the friction tensor of a colloid close to the wall can be decomposed into its bulk value, encoding the particle anisotropy, and a wall-induced correction. This allowed us to formulate a deterministic system of equations which can be solved for arbitrary initial position and orientation of the nano-rod. 

For starting points at a large polar angle $\alpha$, we find that gliding trajectories are governed by the shape anisotropy of the rod, and the wall plays no significant role. However, for smaller approach angles, hydrodynamic interactions with the wall significantly alter the angle at which the rod approaches the near-pore region. We have shown the extent of this region to be as large as $\pi/8$ which accounts for nearly 40\% of spherical area in 3D, thus signifying the importance of wall effects in the proper modelling of dynamics in confined geometry. 

Our work shows, basing on a scaling analysis, that it is justified to neglect the role of Brownian motion in the near-pore region. Previous works on a similar system included Brownian motion on the level of translational motion only, by imposing a constant diffusion coefficient of the particle~\cite{qiao2020capture}. However, in order to properly resolve the question of Brownian displacements and rotations, one should should account for two facts: a) the non-spherical shape of the rod which makes its diffusion anisotropic even in a bulk system, and b) the presence of the wall which renders the anisotropic diffusion tensor of the particle a wall-particle distance-dependent and introduces translation-rotation coupling. Even then, we would expect the effect of Brownian motion to be pronounced outside the near-pore region which is dominated by electrostatic interactions. Including these effects would be an interesting direction of future research.

The analysis of trajectories close to the pore remains a separate issue, since its non-planar geometry has a significant influence on the trajectories, both by changing the structure of the electric field, which can no longer be modelled by a point charge, and by the different character of hydrodynamic interactions, where the pore opening shapes the lubrication flow and friction landscape for the colloid. For an insight into these dynamics, detailed models of the pore structure should be implemented.

% \begin{itemize}
%     \item at large and moderate distances can use approximate analytical schemes
%     \item at close distances, accurate simulations needed (such as \textsc{hydromultipole})
%     \item at close distances, geometric details of the nanopore become important (i.e. 
%     \item we included both sources of anisotropy: anisotropy of the particle itself and of the wall-bounded geometry
%     \item included and assessed both the effect of anisotropic HI on the (a) statics and (b) dynamics of the process in the absence of Brownian motion
%     \item added proper treatment of the Langevin equation for an anisotropic molecule in a confined geometry
%     \item the inclusion of Brownian motion shows ?
    
% \end{itemize}

\section*{Acknowledgments}
 We thank Piotr Szymczak for helpful discussions. The work of RW and ML was supported by the National Science Centre of Poland grant Sonata to ML no. 2018/31/D/ST3/02408.

\appendix
\section{Bulk friction tensors of an ellipsoid}\label{appEllipsoid}

The elements of the bulk friction tensor for an ellipsoidal colloid are known analytically~\cite{KimKarrila}. For a prolate spheroid with a long axis $a=L/2$ and a short axis $c$, corresponding to an eccentricity $e = \sqrt{a^2-c^2}/a$, the bulk friction coefficients in $\bzz_0$ can be written as
\begin{align}
\zeta^{t}_\parallel &= (6\pi \eta a) \frac{8}{3} e^3 \Bigl[-2e + \left(1+e^2\right)\ell\Bigr]^{-1}, \\
\zeta^{t}_\perp &= (6 \pi \eta a) \frac{16}{3} e^3 \Bigl[2e + \left(3e^2-1\right)\ell\Bigr]^{-1}, \\
\zeta^r_\parallel & = (8\pi \eta a^3) \frac{4}{3} e^3(1-e^2) \Bigl[2e - \left(1-e^2\right)\ell\Bigr]^{-1}, \\
\zeta^r_\perp & = (8\pi \eta a^3)  \frac{4}{3} e^3(2-e^2) \Bigl[-2e + \left(1+e^2\right)\ell\Bigr]^{-1}, \\
\zeta^{dr} &=(\pi \eta  a^3) \frac{16}{3}   e^5 \Bigl[-2e + \left(1+e^2\right)\ell\Bigr]^{-1}, \\
\ell &=  \log \left(\frac{1+e}{1-e}\right).
\end{align}
The last coefficient, $\zeta^{dr}$, links the stresslet (symmetric dipole moment) on the spheroid with the rate-of-strain tensor of an external flow.

\section{Wall correction terms}\label{appA}

The correction terms are described in detail in Ref.~\cite{Lisicki2016nearwall}. It is most convenient to specify the components of the near-wall friction tensor in a body-fixed frame of reference. It is defined by a set of basis vectors $\left\{ \bm{u},\bm{u}_{\perp 1},\bm{u}_{\perp 2}\right\}$, where $\bm{u}$ is the director along the long axis of the nano-rod, $\bm{u}_{\perp 1}=(\bm{e}_z\times \bm{u})/\left\vert \bm{e}_z\times \bm{u} \right\vert$ is parallel to the wall and perpendicular to the particle axis, and $\bm{u}_{\perp 2}=\bm{u}_{\perp 1}\times \bm{u}$ completes the orthonormal basis. We can write the tensors in Eqs. \eqref{corrtt}--\eqref{corrrr} explicitly in the body-fixed frame RW.  For convenience, we define the following shorthand notation
\begin{equation}
    \text{s} \equiv \sin \theta, \qquad \text{c} \equiv \cos\theta.
\end{equation}
For the translational part \eqref{corrtt}, we find the correction's angular dependence as
\begin{align}\label{angulartt}  
&\mathbf{A}_1 =-\frac{3}{16\pi\eta}
\begin{pmatrix}
(\zeta^{t}_\parallel)^2(1+\text{c}^2) & 0 & -\zeta^{t}_\parallel\zeta^{t}_\perp \text{s}\,\text{c} \\
0 & (\zeta^{t}_\perp)^2 & 0 \\
-\zeta^{t}_\parallel\zeta^{t}_\perp \text{s}\,\text{c} & 0 & (\zeta^{t}_\perp)^2 (1 + \text{s}^2) 
\end{pmatrix}, \\ 
&\mathbf{A}_2 = \frac{9}{256\pi^2 \eta^2} 
\begin{pmatrix}
A_\parallel & 0 & A_{\parallel\perp} \\
0 &  A_{\perp 1} & 0 \\
A_{\parallel\perp} & 0 & A_{\perp 2}
\end{pmatrix}.
\end{align}
with the coefficients
\begin{align} 
A_{\perp 1} &=(\zeta^{t}_\perp)^3, \\
A_{\parallel} &=(\zeta^{t}_\parallel)^3(1+\text{c}^2)^2 + \zeta^{t}_\perp(\zeta^{t}_\parallel)^2 \text{s}^2\text{c}^2, \\ 
A_{\perp 2} &= (\zeta^{t}_\perp)^3(1+\text{s}^2)^2 + (\zeta^{t}_\perp)^2\zeta^{t}_\parallel \text{s}^2
\text{c}^2, \\ 
A_{\parallel\perp} &=-\zeta^{t}_\parallel\zeta^{t}_\perp[\zeta^{t}_\parallel(1+\text{c}^2)+\zeta^{t}_\perp(1+\text{s}^2)] \text{s}\,\text{c}.
\end{align}
The translation-rotation coupling part reads
\begin{align}
\mathbf{B} =\frac{3\zeta^{dr}}{26\pi\eta}
\begin{pmatrix}
0 & \zeta^{t}_\parallel (1+\text{c}^2) \text{s} & 0 \\
0 & 0 & \zeta^{t}_\perp \text{c} \\
0 & -\zeta^{t}_\perp (1+\text{s}^2)\text{c} & 0 
\end{pmatrix}.
\end{align}
Finally, the rotational part $\mathbf{C}$ in Eq. \eqref{corrrr} has the form
\begin{align} \label{angularrr}
\mathbf{C} =& -\frac{1}{16\pi\eta}\begin{pmatrix}
(\zeta^{r}_\parallel)^2(5-3\text{c}^2) & 0 & 3 \zeta^{r}_\parallel\zeta^{r}_\perp \text{s}\,\text{c} \\
0 & 5(\zeta^{r}_\perp)^2 & 0 \\
3 \zeta^{r}_\parallel\zeta^{r}_\perp \text{s}\,\text{c}  & 0 & (\zeta^{r}_\perp)^2(5 - 3 \text{s}^2)
\end{pmatrix} + \\  \nonumber 
&\frac{3\zeta^{dr}}{16\pi\eta}\begin{pmatrix}
0 & 0 & \zeta^{r}_\parallel \text{s}\,\text{c} \\
0 & -2\zeta^{r}_\perp(1-2 \text{c}^2) & 0 \\
\zeta^{r}_\parallel \text{s}\,\text{c}  & 0 & 2 \zeta^{r}_\perp \text{c}^2
\end{pmatrix} + \\ \nonumber
&-\frac{3(\zeta^{dr})^2}{16\pi\eta}\begin{pmatrix}
0 & 0 & 0 \\
0 & 3+\text{c}^2-\text{c}^4 & 0 \\
0  & 0 & 1+2\text{c}^2
\end{pmatrix}.
\end{align}

%%%% END OF COPIED IN

\section*{References}

\bibliographystyle{unsrt}
\bibliography{nanopore_rod}

\end{document}